\documentclass[aps,preprint]{revtex4}%
\usepackage{amsfonts}
\usepackage{amsmath}
\usepackage{amssymb}
\usepackage{subfigure}
\usepackage{graphicx}
\usepackage{color}%
\usepackage{makecell}

\usepackage{cleveref}

\begin{document}
	\title{Thermodynamic instability and the violation of reverse isoperimetric inequality in Torus-like black hole}
	\author{}
	
\author{Yue Song$^{a,b}$}
\email{syphysics@std.uestc.edu.cn}
\author{Yiqian He$^{a,b}$}
\email{heyiqian@std.uestc.edu.cn}
\author{Benrong Mu$^{a,b,c}$}
\email{benrongmu@cdutcm.edu.cn}
\affiliation{$^{a}$ Center for Joint Quantum Studies, College of Medical Technology,
Chengdu University of Traditional Chinese Medicine, Chengdu, 611137, PR China}
\affiliation{$^{b}$ School of Physics, University of Electronic Science and Technology of China, Chengdu, 611731, China}
\affiliation{$^{c}$Center for Theoretical Physics, College of Physics,
Sichuan University, Chengdu, 610064, PR China}

	\begin{abstract}
		Recently, conjectures have been made that there is a correlation between thermodynamic instability and super-entropy~\cite{{Johnson:2019mdp},{Cong:2019bud}}. In this paper, we started with torus-like black holes and verified the correlation mentioned above. After solving the specific analytical solutions of the specific heat capacity at constant volume $C_{V}$, the specific heat capacity at constant pressure $C_{P}$ and plotting the figures, we obtained the results which are consistent with the hypothesis from Ref. \cite{Cong:2019bud}. To further verify the generality of conjectures, we explored the BTZ black hole in nonlinear electrodynamics and phantom AdS black hole. And then the results are summarized via a table. Ultimately, the comparison among these three black holes leads us to the conclusion that there is no essential connection between thermodynamic instability and super-entropy.

	\end{abstract}
	
	\maketitle
	\tableofcontents{}
	
	\bigskip{}
	
	

	\section{Introduction}
	Black hole as a classical thermodynamic system with plentiful intriguing characteristics has received much attention constantly. A black hole in standard framework is allotted to the temperature $T$, the entropy $S=\frac{A}{4}$ and the energy $U=M$, which is limited by the first law of thermodynamics $dU=TdS$. In reality, the gravitational system is not always stable. Notably, a classical case is contributed by the Schwarzschild black hole. Considering the Hawking radiation, the black hole loses part mass-energy, which leads to the higher temperature as well as the more radiation, and hence the quicker mass loss. Since then, scientists have observed a great deal of black holes and summarized the quantum instability of that can be judged by the sign of the specific heat
	\begin{equation}\label{eqn:Q15}
		C=T\frac{\partial S}{\partial T}=-\frac{1}{8\pi T^{2}}. \\
	\end{equation}
	
	When the specific heat capacity is less than zero, the black hole is thermodynamically
	unstable.
	
	The conjecture proposed in Ref.~\cite{Cong:2019bud} that super-entropy black holes are associated with thermodynamic instability, is an exploration of the physical meaning behind super-entropy. Until now, there has been no reasonable explanation for the existence of super-entropy black holes. It would be of great significance if this conjecture could be confirmed. Hence, we have decided to test our conclusion on a number of black holes and then focused on the methods of that.
	Super-entropy black holes described above originate from the violation of inverse isoperimetric inequality~\cite{Cvetic:2010jb}. When a black hole satisfies the corresponding ratio $\mathcal{R}<1$, we call it a super-entropy black hole, and $\mathcal{R}$ is calculated as follows
	\begin{equation}\label{eqn:Q11}
		\mathcal{R}=(\frac{(d-1)V}{\omega_{d-2}})^{\frac{1}{d-1}}(\frac{\omega_{d-2}}{A}),\\
	\end{equation}
where $\omega_{d}=\frac{2\pi^{\frac{d+1}{2}}}{\Gamma(\frac{d+1}{2})}$.

	In above equations, $d$ refers to the dimension of the black hole, while $V$ and $A$ refer to the volume and surface area of the black hole respectively.
	We utilize the result of Eq. $\left(\ref{eqn:Q11}\right)$ to determine whether the black hole is a super-entropy black hole or not. Until now, massive super-entropy black holes have been investigated in Refs.~\cite{{Hennigar:2014cfa},{Klemm:2014rda},{Hennigar:2015cja},{Brenna:2015pqa},{Song:2023kvq},{He:2023tiz}}.
	
	It is remarkable that for a black hole in the extended phase space~\cite{{Kubiznak:2016qmn},{Dolan:2010ha},{Caldarelli:1999xj}}, the mass should be explained as the entropy~\cite{Kastor:2009wy}. Meanwhile the cosmological constant $\Lambda
	$ is related to the pressure $P$ as $P=-\frac{1}{8\pi}\Lambda$~\cite{{Kastor:2009wy},{Dolan:2011xt},{Dolan:2014jva}}, and the volume $V=(\frac{\partial M}{\partial P})_{S}$ is interpreted as the conjugate variable of $P$. Based on these alterations, the first law of thermodynamic will change to $dH=TdS+VdP$.
	
	Generally, the specific heat capacity of a black hole, which measures the thermodynamic stability of the system, is $C_{P}$ habitually (the specific heat capacity that guarantees the invariance of the cosmological constant). However, in the extended phase space, the volume $V$ is introduced, so the $C_{V}$ naturally needs to be included in the test.

In Ref.~\cite{Johnson:2019mdp}, Clifford et al. proposed the super-entropy black hole always has negative $C_{V}$. Afterward Wan et al. found a counterexample and put forward a further conjecture that for a super-entropy black hole, where $C_{V}>0$, the specific heat at constant pressure $C_{P}$ is negative, which leads to the black hole investigated is thermodynamically unstable ~\cite{Cong:2019bud}.

	The torus-like black hole is the static solution of the Einstein-Maxwell equation, of which the event horizon possesses $S_{1}\times S_{1}\times\mathbb{R}$ topology~\cite{{Huang:1995zb},{Lemos:1994xp},{Hong:2020zcf},{Lemos:1995cm},{Han:2019kjr},{Feng:2021vey}}. Particularly, every surface of the black hole in the space-time with a toroidal topology at fixed radius, differs from that in the asymptotically flat space time. The distinct structure probably brings the unique to the black hole, which inspires our interest in it. Hence, we analyzed the thermodynamic stability of the black hole and compared the result with other typical black holes, BTZ black hole in nonlinear electrodynamics and the Phantom AdS black hole.

The rest of paper is organized as follows. In section~\ref{sec:A}, we reviewed the thermodynamic
properties in the torus-like black hole. In section~\ref{sec:B}, we analyzed the thermodynamic instability of torus-like black hole and plotted the curves of specific heat capacity of that. Finally, we verified the conjecture in plenty of black holes and summarized our results in section~\ref{sec:C}.
	
	\section{Thermodynamics in Torus-like black hole }
\label{sec:A}

	The metric of the torus-like black hole mentioned in Ref.~\cite{Feng:2021vey} is conveyed via
	\begin{equation}\label{eqn:Q1}
		ds^{2}=-f(r)dt^{2}+f^{-1}(r)dr^{2}+r^{2}(d\theta^{2}+d\psi^{2}), \\
	\end{equation}
	where
	\begin{equation}\label{eqn:Q2}
		f(r)=-\frac{\Lambda r^{2}}{3}-\frac{2M}{\pi r}+\frac{4Q^{2}}{\pi r^{2}}.\\
	\end{equation}
	
	As $\Lambda<0$, there are coordinate singularities at the radii of the event horizon. Meanwhile, the radii of event horizon $r_{\pm}$ satisfy
	\begin{equation}\label{eqn:Q4}
		-\frac{\Lambda r_{\pm}^{2}}{3}-\frac{2M}{\pi r_{\pm}}+\frac{4Q^{2}}{\pi r_{\pm}^{2}}=0,\\
	\end{equation}
	where $r_{-}$ and $r_{+}$ are the radii of the inner and outer event horizons of the black hole respectively.
	
	In the extended phase space, the cosmological constant is associated with thermodynamic pressure and the corresponding conjugate quantity of cosmological constant is regarded as thermodynamic volume i.e
	\begin{equation}\label{eqn:Q5}
		P=-\frac{\Lambda}{8\pi},\\
		V=\frac{4\pi^{2}r_{h}^{3}}{3},\\
	\end{equation}
	where $r_{h}$ denotes the radius of horizon.
	
	Utilizing $T=\frac{1}{4\pi}\frac{\partial f}{\partial r}$, the temperature of a torus-like black hole can be obtained by
	\begin{equation}\label{eqn:Q7}
		T=\frac{-12Q^{2}+3Mr_{h}-\pi r_{h}^{4}\Lambda}{6\pi^{2}r_{h}^{3}}.\\
	\end{equation}
	
	And the entropy of black hole is expressed as
	\begin{equation}\label{eqn:Q8}
		S=\pi^{2}r_{h}^{2}.\\
	\end{equation}
	
	Via solving $f(r)=0$, the expression for mass can be calculated as follows
	\begin{equation}\label{eqn:Q10}
		M=\frac{-12Q^{2}-\pi r_{h}^{4}\Lambda}{6r_{h}}.\\
	\end{equation}

	Initially, it is essential to verify whether a torus-like black hole is a super-entropy black hole. As explained above, the super-entropy black holes can be judged via the violation of inverse isoperimetric inequality i.e. $\mathcal{R}<1$. The ratio $\mathcal{R}$ for four dimension is calculated by the following formula
	\begin{equation}\label{eqn:Q103}
	\mathcal{R}=(\frac{3V}{4\pi})^{\frac{1}{3}}(\frac{4\pi}{\mathcal{A}})^{\frac{1}{2}},\\
	\end{equation}
    where $\mathcal{A}=4\pi^{2}r_{h}^{2}$.

	After substituting the variables associated with torus-like black holes, we obtained
	\begin{equation}\label{eqn:Q104}
	\mathcal{R}=(\frac{4\pi^{2}r_{h}^{3}}{4\pi})^{\frac{1}{3}}(\frac{4\pi}{4\pi^{2}r_{h}^{2}})^{\frac{1}{2}}=\pi^{-\frac{1}{6}}<1.
    \end{equation}

	It is obvious that $\mathcal{R}<1$, which indicates that torus-like black holes are super-entropy black holes.
	
\section{Verifying the relation between super-entropy and thermodynamic instability in Torus-like black hole }
	\label{sec:B}
	From the verification procedure given in Ref.~\cite{Johnson:2019mdp}, we learned that the figures of $C_{V}$ and $C_{P}$ related to the radius $r_{+}$ can be used for checking thermodynamic instability.
	The formula for $C_{P}$ is as follows
	\begin{equation}\label{eqn:Q13}
		C_{P}=T\frac{\partial S}{\partial T}\mid_{P}. \\
	\end{equation}

	The calculation of $C_{V}$ is based on the $C_{P}$ and obtained via the following expression
	\begin{equation}\label{eqn:Q14}
		C_{V}=C_{P}-TV\alpha_{P}^{2}\kappa_{T}, \\
	\end{equation}
	where $\alpha_{P}\equiv V^{-1}\frac{\partial V}{\partial T}\mid_{P}$, $\kappa_{T}\equiv-V\frac{\partial P}{\partial V}\mid_{T}$.

	If the $C_{P}$, $C_{V}$ of a black hole is greater than zero at the same time, this black hole is thermodynamic stable. Otherwise, it is situated in thermodynamic instability.

    Utilizing the above formulas, we can acquire the specific analytical solutions of $C_{V}$ and $C_{P}$ for torus-like black holes
		\begin{equation}\label{eqn:Q101}
	C_{P}=2\pi^{2}r^{2}(1-\frac{4Q^{2}}{3Q^{2}+2P^{2}r^{4}}), \\
	\end{equation}
	
		\begin{equation}\label{eqn:Q102}
	C_{V}=-\frac{\pi^{2}(-Q^{2}+2P^{2}r^{4})(-2P^{2}r^{2}+\pi^{2}\sqrt{P^{2}r^{4}})}{3P^{2}Q^{2}+2P^{4}r^{4}}. \\
	\end{equation}

    On the basis mentioned above, we draw the figures of $C_{V}$ curve and $C_{P}$ curve under different quantity of electric charge $Q$ to test the correctness of conjecture proposed by Ref.~\cite{Cong:2019bud}.

\begin{figure}
		\begin{center}
			\subfigure[{}]{
				\includegraphics[width=0.48\textwidth]{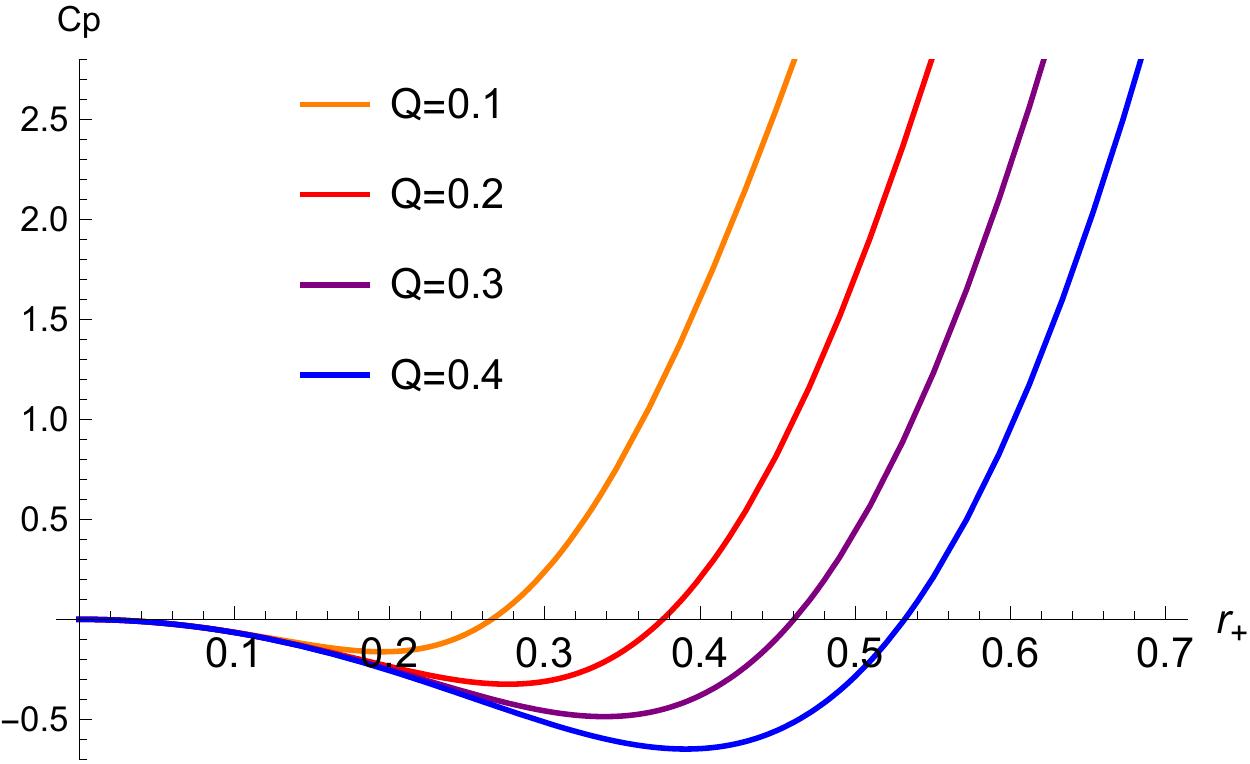}\label{fig:trp}}
			\subfigure[{}]{
				\includegraphics[width=0.48\textwidth]{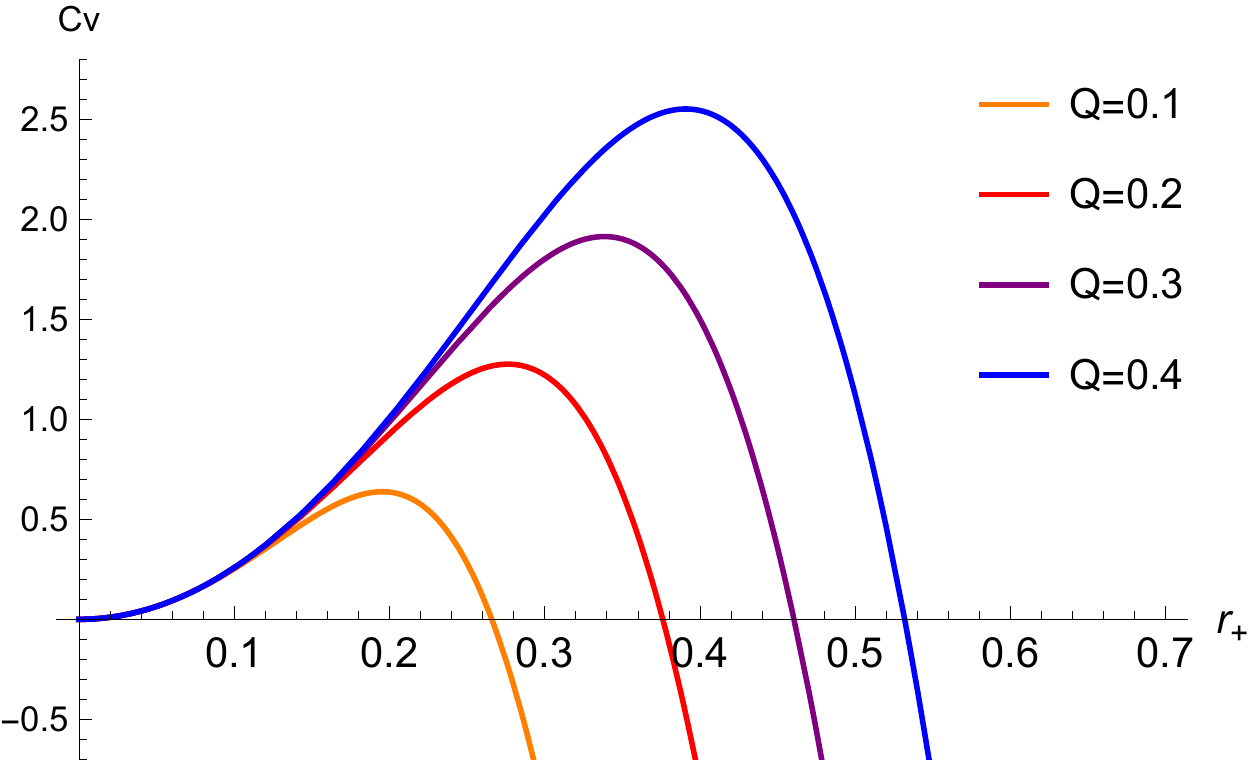}\label{fig:trv}}
		\end{center}
		\caption{The curves of specific heat capacity plotted with $Q=0.1$, $0.2$, $0.3$ and $0.4$, where $P=1$  }%
		\label{fig:C01}
	\end{figure}

    Carefully observing Fig.\ref{fig:C01}, it is obvious that when $C_{V}$ is less than 0, $C_{P}$ must be greater than 0. Meanwhile, when $C_{V}$ is greater than 0, $C_{P}$ must be less than 0. $C_{V}$ and $C_{P}$ cannot be positive at the same time, which means that torus-like black holes are thermodynamic unstable. This verifies the conclusion proposed by Wan et al. that super-entropy denotes thermodynamic instability.

\section{Discussion and Conclusion}	
	\label{sec:C}
 After investigating the hypotheses of Refs.~\cite{Johnson:2019mdp,Cong:2019bud}, we set out to verify the validity of those via a torus-like black hole. Above all we reviewed the thermodynamic properties of the torus-like black hole and confirmed it as super-entropy black hole. Then the specific analytical solutions of the $C_{V}$, $C_{P}$ for the torus-like black hole are calculated and the corresponding curves are drawn. With the Fig.\ref{fig:C01}, we obtained that the torus-like black hole satisfies the conjecture presented by Wan et al.

 To further prove the correctness of the conjectures, we explored the BTZ black hole in nonlinear electrodynamics and phantom AdS black hole (the exact steps are placed in the Appendix) and concluded the Table \ref{tab:01}. By analysing the results in the table, we found that super-entropy is not necessarily associated with thermodynamic stability. For the BTZ black hole in nonlinear electrodynamics, it contains thermodynamic stable regions while belonging to the super-entropy black holes. In contrast, the phantom AdS black hole satisfies thermodynamic instability but is not a super-entropy black hole. At this point, we have used the three examples to demonstrate that super-entropy is not essentially related to thermodynamic instability. Thus the conjectures are not applicable for all the black holes.

\begin{table}
	\centering
	\caption{Analysis of the correlation between super-entropy and thermodynamic stability}
	\begin{tabular}{|l|l|l|}
		\hline
		\multicolumn{1}{|c|}{Types of black holes}
		& The magnitude of $\mathcal{R}$ versus 1  & \thead{Is there a region where \\the $C_{V}$, $C_{P}$ is greater than zero simultaneously}  \\
		\hline
		 torus-like black hole& $\mathcal{R}<1$ &  not exist  \\
		\hline
		BTZ black hole& $\mathcal{R}<1$ &  exist \\
		\hline
		phantom AdS black hole& incomparable &  exist \\
		\hline
	\end{tabular}
\label{tab:01}
\end{table}

\begin{acknowledgments}
	We are grateful to Deyou Chen, Rui Yin, Jing Liang, Peng Wang, Haitang Yang, Jun Tao and Xiaobo Guo for useful discussions. This work is supported in part by NSFC (Grant No. 11747171), Xinglin Scholars Project of Chengdu University of Traditional Chinese Medicine (Grant no.QNXZ2018050).
\end{acknowledgments}

\section{APPENDIX}

	\subsection{APPENDIX: thermodynamic instability in BTZ black hole in nonlinear electrodynamics}
\label{sec:D}
	Since the notable BTZ black hole~\cite{{Banados:1992wn},{Banados:1992gq}} were investigated, the BTZ black hole in (2 + 1) dimension has attracted plenty of physicists to explore. In (2 + 1) dimensional space-time, the static and spherically symmetric metric is considered as
	\begin{equation}\label{eqn:Q31}
		ds^{2}=-f(r)dt^{2}+\frac{dr^{2}}{f(r)}+r^{2}d\phi^{2}.
	\end{equation}

	According to Ref.~\cite{Paul:2023qvh}, the solution of BTZ black hole within the framework of nonlinear electrodynamics is
	\begin{equation}\label{eqn:Q32}
		f(r)=-M+\frac{r^{2}}{l^{2}}-\frac{2Qr}{a}-\frac{2r^{2}}{a^{2}}+\frac{(Qa+2r)\sqrt{Qar+r^{2}}}{a^{2}}-\frac{Q^{2}}{2}ln[\frac{e(\frac{Qa}{2}+r+\sqrt{Qar+r^{2}})}{2l}].
	\end{equation}

   The elementary thermodynamic characters can be calculated via the same procedures mentioned above
	\begin{equation}\label{eqn:Q33}
		\begin{aligned}
			&S=\frac{\pi r}{2},\\
			&T=\frac{1}{4\pi}(\frac{2r}{l^{2}}-\frac{4r}{a^{2}}-\frac{2Q}{a}+\frac{4\sqrt{r}\sqrt{Qar}}{a^{2}}),\\
			&V=\pi r^{2}-\frac{Q^{2}}{32P},\\
			&P=\frac{1}{8\pi l^{2}}.\\
		\end{aligned}
	\end{equation}

    As for the verification of super-entropy, via using Eq. {$\left(\ref{eqn:Q11}\right)$}, we can obtain the result for the ratio $\mathcal{R}$
    \begin{equation}\label{eqn:Q106}
    \mathcal{R}=(1-\frac{Q^{2}}{32\pi Pr^{2}})^{\frac{1}{2}}<1,
    \end{equation}
which denotes the BTZ black holes in nonlinear electrodynamics are super-entropy black holes.

	Similarly, using Eq. {$\left(\ref{eqn:Q13}\right)$} and Eq. {$\left(\ref{eqn:Q14}\right)$}, the $C_{P}$ and $C_{V}$ are given by
	
	\begin{equation}\label{eqn:Q105}
	C_{P}=-\frac{\pi(Qa+2r-8a^{2}P\pi r-2\sqrt{r}\sqrt{aQ+r})}{2(-2+8a^{2}P\pi+\frac{aQ+2r}{\sqrt{r}\sqrt{aQ+r}})},
	\end{equation}

	\begin{equation}\label{eqn:Q34}
		C_{V}=\frac{a^{2}\pi(-1+8P^{3}\pi)r(aQ+r)(-aQ-2r+8a^{2}P\pi r+2\sqrt{r}\sqrt{aQ+r})}{2P^{2}(2r-2\sqrt{r}\sqrt{aQ+r}+a(Q+8aP\pi\sqrt{r}\sqrt{aQ+r}))^{2}}.
	\end{equation}
		\begin{figure}
		\begin{center}
			\subfigure[{}]{
				\includegraphics[width=0.48\textwidth]{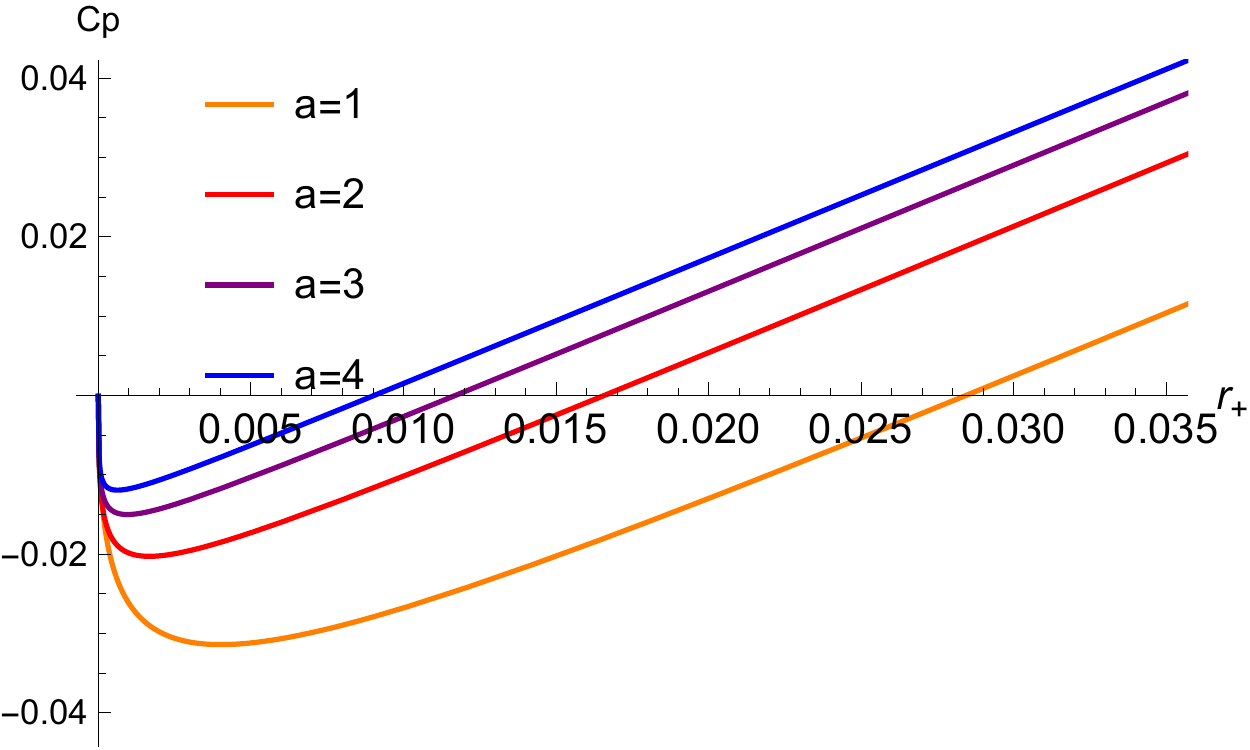}\label{fig:BTZcpa}}
			\subfigure[{}]{
				\includegraphics[width=0.48\textwidth]{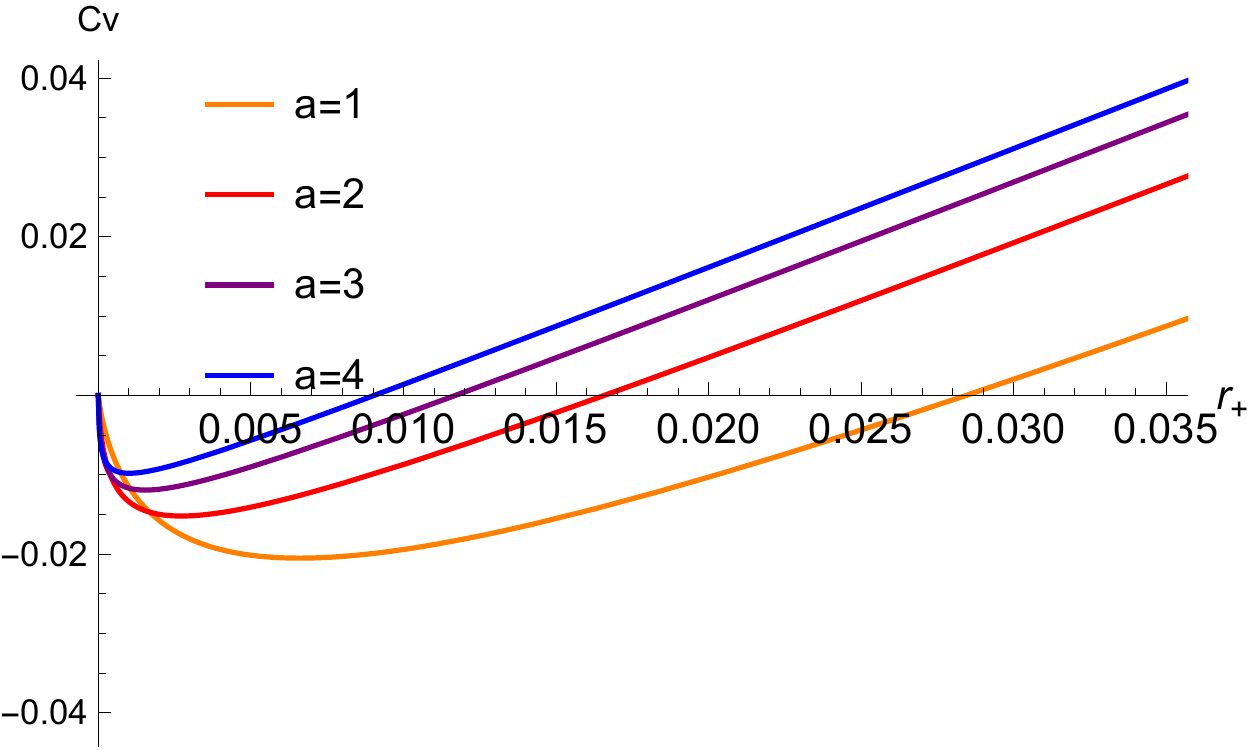}\label{fig:BTZcva}}
		\end{center}
		\caption{The curves of specific heat capacity plotted with $a=1$, $2$, $3$ and $4$ in BTZ black holes in nonlinear electrodynamics, where $P=1$  }%
		\label{fig:C02}
	\end{figure}

	\begin{figure}
		\begin{center}
			\subfigure[{}]{
				\includegraphics[width=0.48\textwidth]{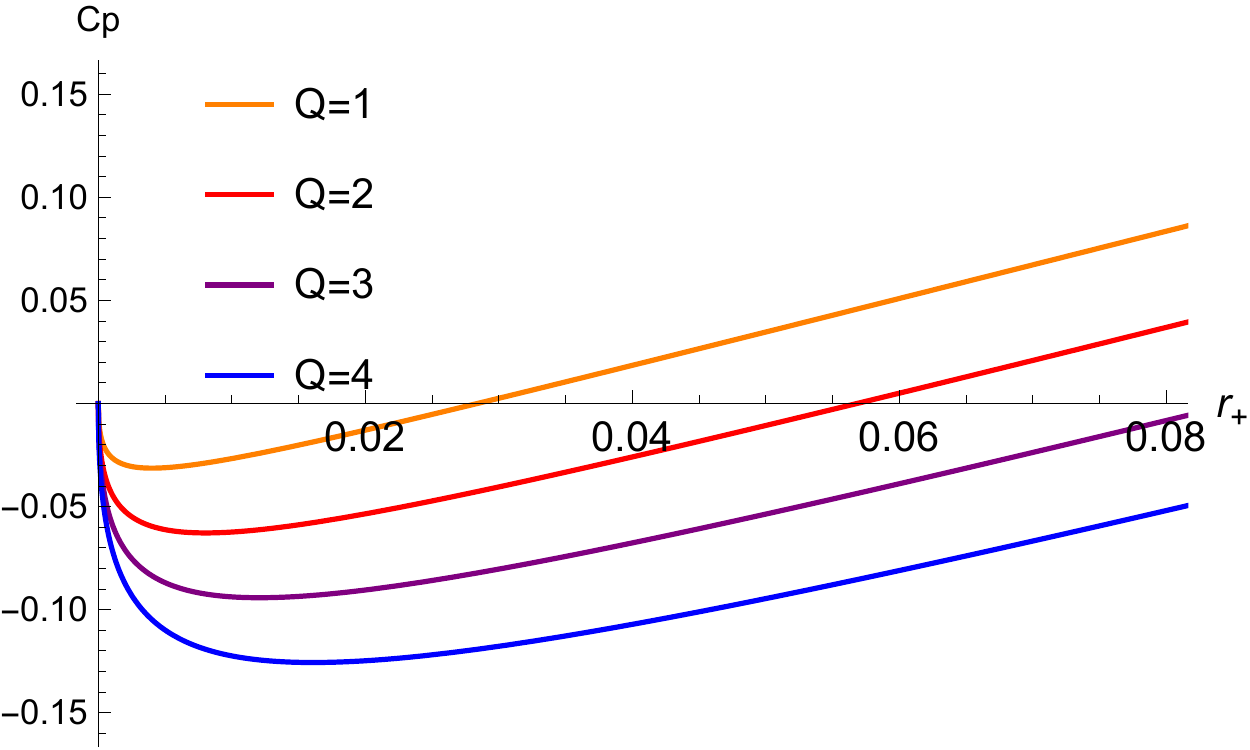}\label{fig:BTZcpQ}}
			\subfigure[{}]{
				\includegraphics[width=0.48\textwidth]{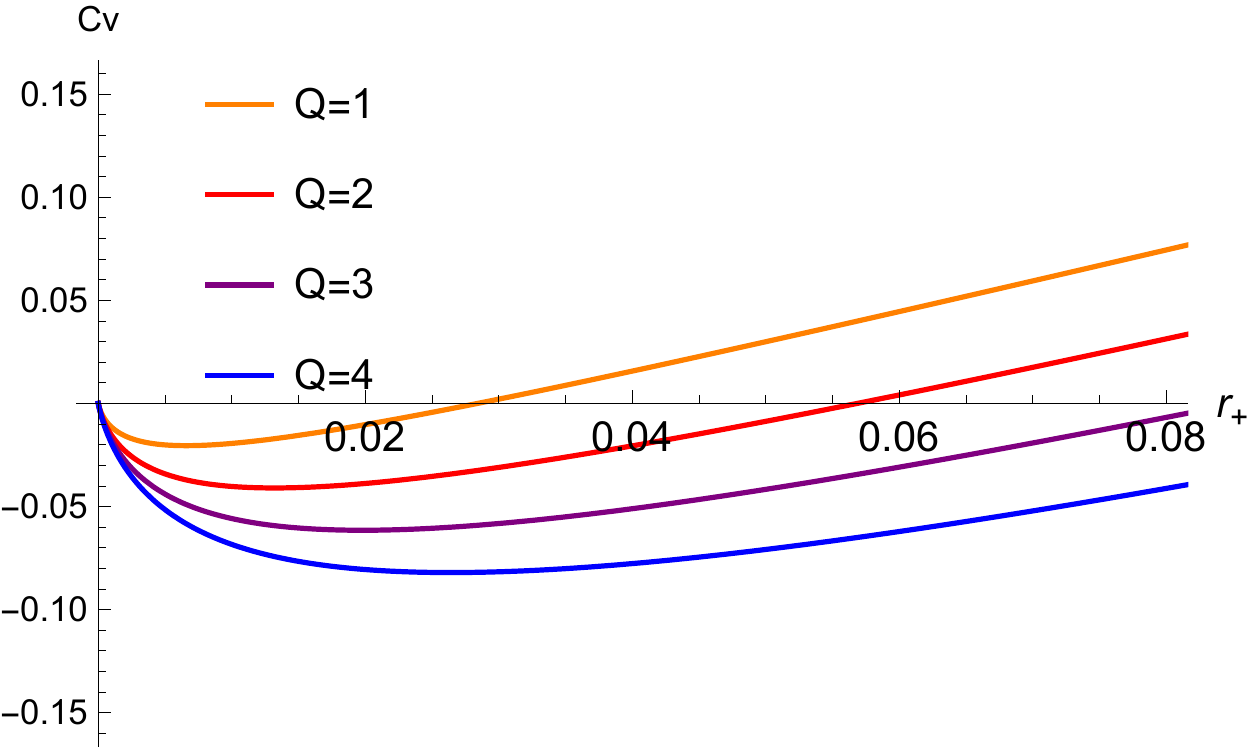}\label{fig:BTZcvQ}}
		\end{center}
		\caption{The curves of specific heat capacity plotted with $Q=1$, $2$, $3$ and $4$ in BTZ black holes in nonlinear electrodynamics, where $P=1$  }%
		\label{fig:C03}
	\end{figure}

	Fig.\ref{fig:C02} and Fig.\ref{fig:C03} show that $C_{V}$ and $C_{P}$ maintain the identical sign, which implies the existence of a thermodynamic stable region.

	\subsection{APPENDIX: thermodynamic instability in phantom AdS black hole}	
\label{sec:E}
	Under the geometric units, the Einstein-Hilbert action with cosmological constant $\Lambda$ coupled to the electromagnetic field~\cite{Mo:2018hav} can be expressed as
	\begin{equation}\label{eqn:Q21}
		S=\int d^{4}x\sqrt{-g}(R+2\eta F_{\mu\nu}F^{\mu\nu}+2\Lambda),
	\end{equation}
	where the constant $\eta$ denotes the character of the electromagnetic field. In particular, $\eta=1$ and $\eta=-1$ lead to the classical Maxwell field and the phantom field, respectively. For the latter one, namely the phantom field, the energy density is negative.
	
	In the case of static and spherically space-time, the solution of the action is given by ~\cite{Quevedo:2016cge}
	\begin{equation}\label{eqn:Q22}
		\begin{aligned}
			&ds^{2}=f(r)dt^{2}-\frac{dr^{2}}{f(r)}-r^{2}(d\theta^{2}+sin^{2}\theta d\phi^{2}),\\
			&f(r)=1-\frac{2M}{r}-\frac{\Lambda}{3}r^{2}+\eta\frac{q^{2}}{r^{2}}.\\
		\end{aligned}
	\end{equation}
	
	Likewise, the thermodynamic quantities are obtained in Ref.~\cite{Jardim:2012se}
	\begin{equation}\label{eqn:Q23}
		\begin{aligned}
			&S=\pi r^{2},\\
			&T=\frac{1}{4S^{\frac{3}{2}}l^{2}}(Sl+3S^{2}-\eta q^{2}l^{2}),\\
			&V=\frac{4\pi}{3}r^{3},\\
			&P=\frac{3}{8\pi l^{2}}.\\
		\end{aligned}
	\end{equation}
	
	 As for the verification of super-entropy, via using Eq. {$\left(\ref{eqn:Q11}\right)$}, we can obtain the result for the ratio $\mathcal{R}$
	\begin{equation}\label{eqn:Q107}
		\mathcal{R}=((\frac{32}{3})^{\frac{1}{3}}4^{-\frac{1}{4}}\pi^{\frac{7}{12}})Pr^{-\frac{1}{2}},
	\end{equation}
	
	It is difficult to judge whether $\mathcal{R}<1$. Therefore, the phantom AdS black holes are not super-entropy black holes.
	
	Utilizing the Eq. {$\left(\ref{eqn:Q13}\right)$} and Eq. {$\left(\ref{eqn:Q14}\right)$}, the $C_{P}$ and $C_{V}$ are given by
	\begin{equation}\label{eqn:Q24}
	C_{P}=\frac{2\pi r^{2}(-3\eta q^{2}+\frac{2\sqrt{6}\pi^{\frac{3}{2}}r^{2}}{\sqrt{\frac{1}{P}}}+24P\pi^{3}r^{4})}{9\eta q^{2}-\frac{2\sqrt{6}\pi^{\frac{3}{2}}r^{2}}{\sqrt{\frac{1}{P}}}+24P\pi^{3}r^{4}},
	\end{equation}
	\begin{equation}\label{eqn:Q25}
		\begin{aligned}
		&C_{V}=\frac{2\pi r^{2}(-3\eta q^{2}+\frac{2\sqrt{6}\pi^{\frac{3}{2}}r^{2}}{\sqrt{\frac{1}{P}}}+24P\pi^{3}r^{4})}{9\eta q^{2}-\frac{2\sqrt{6}\pi^{\frac{3}{2}}r^{2}}{\sqrt{\frac{1}{P}}}+24P\pi^{3}r^{4}}+\\
		&(8P^{4}(\frac{2\sqrt{6}}{\sqrt{\frac{1}{P}}}-\frac{3\eta q^{2}}{\pi^{\frac{3}{2}}r^{2}}+24P\pi^{\frac{3}{2}}r^{2})(3\eta q^{2}r^{4}-2(1+6\eta q^{2})r^{4}-\frac{2\sqrt{6}\pi^{\frac{3}{2}}r^{6}}{\sqrt{\frac{1}{P}}}\\
		&-24P\pi^{3}r^{8}+\frac{-2-9\eta q^{2}-\frac{10\sqrt{6}\pi^{\frac{3}{2}}r^{2}}{\sqrt{\frac{1}{P}}}-120P\pi^{3}r^{4}}{\sqrt{\frac{1+\frac{8\sqrt{6}\pi^{\frac{3}{2}}r^{2}}{\sqrt{\frac{1}{P}}}+96P\pi^{3}r^{4}}{r^{8}}}}))\\
		&/9\pi^{2}(r^{2})^{\frac{3}{2}}(3\eta P^{2}q^{2}+r(-P+3r))^{2})
		.\\
		\end{aligned}
	\end{equation}

		\begin{figure}
	\begin{center}
		\subfigure[{}]{
			\includegraphics[width=0.48\textwidth]{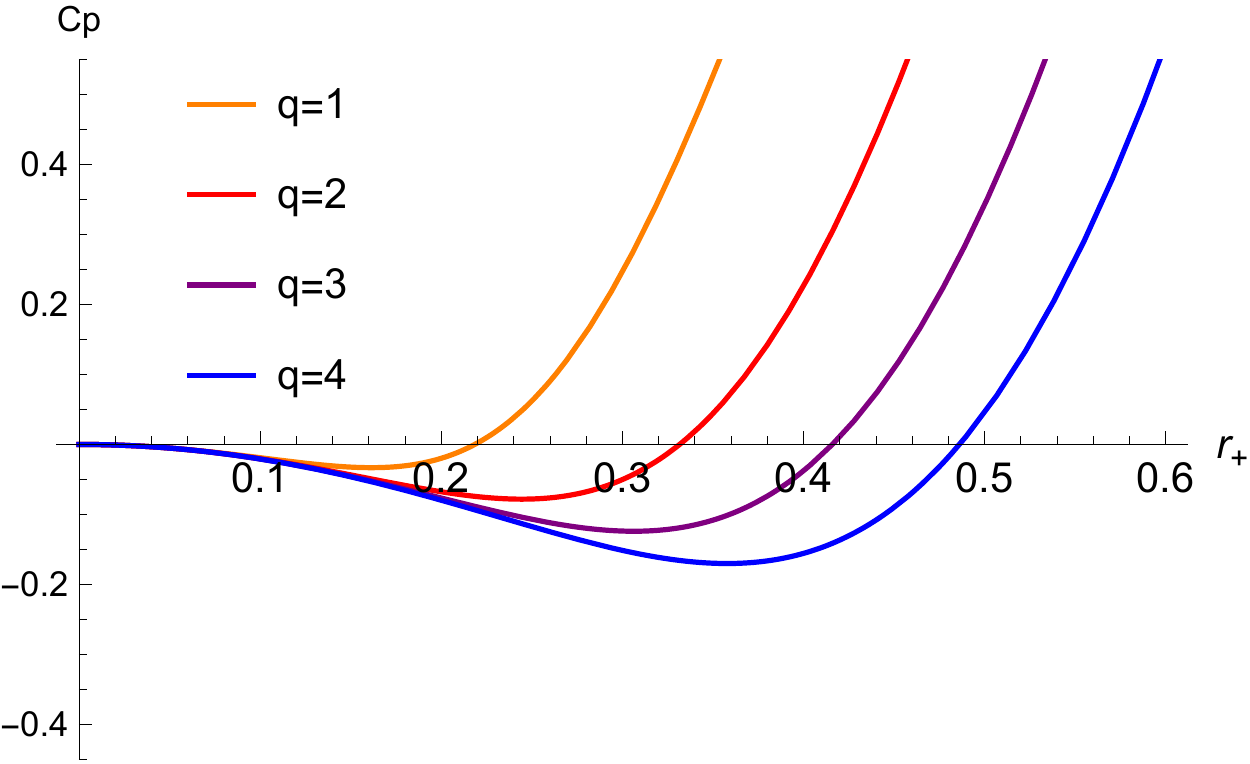}\label{fig:hcpn1}}
		\subfigure[{}]{
			\includegraphics[width=0.48\textwidth]{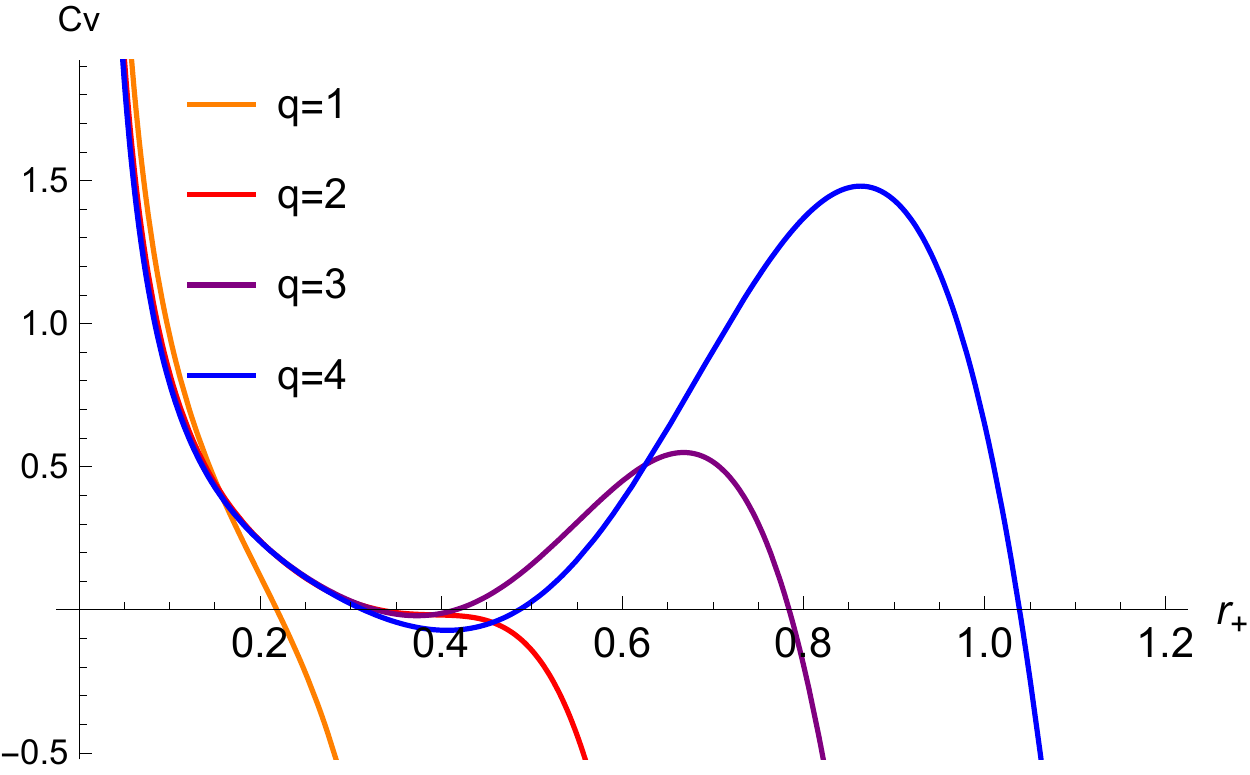}\label{fig:hcvn1}}
	\end{center}
	\caption{The curves of specific heat capacity plotted with $q=1$, $2$, $3$ and $4$ in phantom AdS black holes, where $P=1$, $\eta=1$  }%
	\label{fig:C04}
\end{figure}

		\begin{figure}
	\begin{center}
		\subfigure[{}]{
			\includegraphics[width=0.48\textwidth]{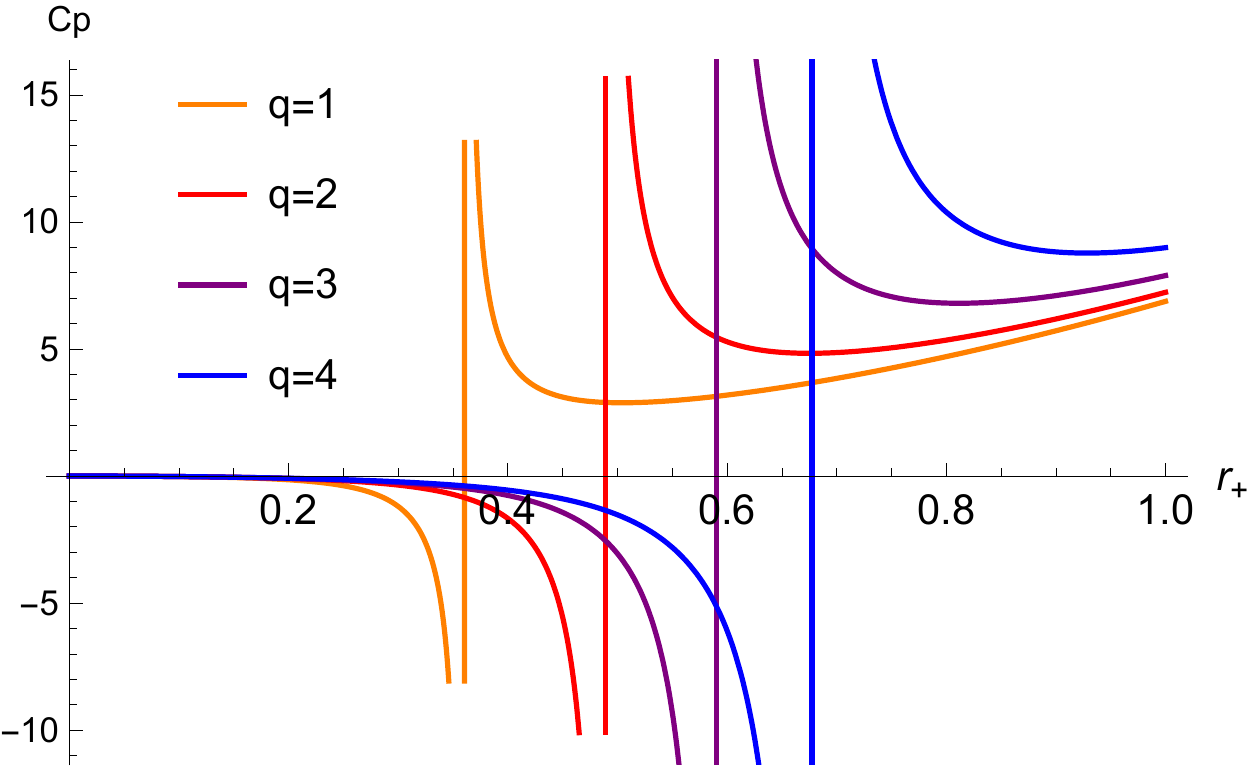}\label{fig:hcpnm1}}
		\subfigure[{}]{
			\includegraphics[width=0.48\textwidth]{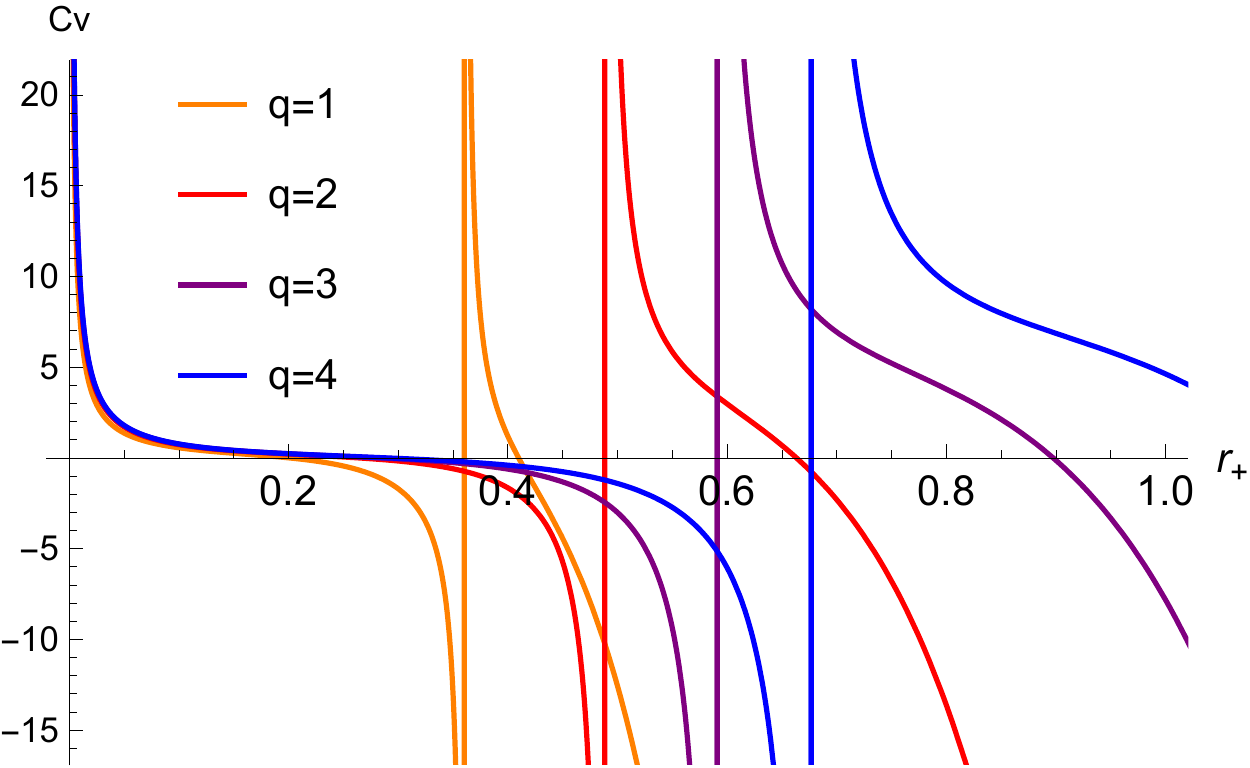}\label{fig:hcvnm1}}
	\end{center}
	\caption{The curves of specific heat capacity plotted with $q=1$, $2$, $3$ and $4$ in phantom AdS black holes, where $P=1$, $\eta=-1$  }%
	\label{fig:C05}
\end{figure}

It can be obviously seen from Fig.\ref{fig:C04} and Fig.\ref{fig:C05} that there is a region where $C_{V}$ and $C_{P}$ are positive concurrently in both figures, which indicates the existence of a region satisfying thermodynamic stability in phantom AdS black holes.


\begin{thebibliography}{999}

\bibitem{Johnson:2019mdp}
C.~V.~Johnson,
``Instability of super-entropic black holes in extended thermodynamics,''
Mod. Phys. Lett. A \textbf{35}, no.13, 2050098 (2020)
doi:10.1142/S0217732320500984
[arXiv:1906.00993 [hep-th]].
\bibitem{Cong:2019bud}
W.~Cong and R.~B.~Mann,
``Thermodynamic Instabilities of Generalized Exotic BTZ Black Holes,''
JHEP \textbf{11}, 004 (2019)
doi:10.1007/JHEP11(2019)004
[arXiv:1908.01254 [gr-qc]].
\bibitem{Cvetic:2010jb}
M.~Cvetic, G.~W.~Gibbons, D.~Kubiznak and C.~N.~Pope,
``Black Hole Enthalpy and an Entropy Inequality for the Thermodynamic Volume,'' Phys. Rev. D \textbf{84}, 024037 (2011)
doi:10.1103/PhysRevD.84.024037 [arXiv:1012.2888 [hep-th]].
\bibitem{Hennigar:2014cfa}
R.~A.~Hennigar, D.~Kubiz\v{n}\'ak and R.~B.~Mann,
``Entropy Inequality Violations from Ultraspinning Black Holes,''
Phys. Rev. Lett. \textbf{115}, no.3, 031101 (2015)
doi:10.1103/PhysRevLett.115.031101
[arXiv:1411.4309 [hep-th]].

\bibitem{Klemm:2014rda}
D.~Klemm,
``Four-dimensional black holes with unusual horizons,''
Phys. Rev. D \textbf{89}, no.8, 084007 (2014)
doi:10.1103/PhysRevD.89.084007
[arXiv:1401.3107 [hep-th]].

\bibitem{Hennigar:2015cja}
R.~A.~Hennigar, D.~Kubiz\v{n}\'ak, R.~B.~Mann and N.~Musoke,
``Ultraspinning limits and super-entropic black holes,''
JHEP \textbf{06}, 096 (2015)
doi:10.1007/JHEP06(2015)096
[arXiv:1504.07529 [hep-th]].

\bibitem{Brenna:2015pqa}
W.~G.~Brenna, R.~B.~Mann and M.~Park,
``Mass and Thermodynamic Volume in Lifshitz Spacetimes,''
Phys. Rev. D \textbf{92}, no.4, 044015 (2015)
doi:10.1103/PhysRevD.92.044015
[arXiv:1505.06331 [hep-th]].

\bibitem{Song:2023kvq}
Y.~Song and B.~Mu,
``Thermodynamic Instabilities of Conformal Gravity Holography in Four Dimensions,''
[arXiv:2304.09760 [gr-qc]].

\bibitem{He:2023tiz}
Y.~He and B.~Mu,
``Thermodynamic instabilities of Kerr-Newman Ads black holes,''
[arXiv:2305.02196 [gr-qc]].



\bibitem{Kubiznak:2016qmn}
D.~Kubiznak, R.~B.~Mann and M.~Teo,
``Black hole chemistry: thermodynamics with Lambda,''
Class. Quant. Grav. \textbf{34}, no.6, 063001 (2017)
doi:10.1088/1361-6382/aa5c69
[arXiv:1608.06147 [hep-th]].

\bibitem{Dolan:2010ha}
B.~P.~Dolan,
``The cosmological constant and the black hole equation of state,''
Class. Quant. Grav. \textbf{28}, 125020 (2011)
doi:10.1088/0264-9381/28/12/125020
[arXiv:1008.5023 [gr-qc]].

\bibitem{Caldarelli:1999xj}
M.~M.~Caldarelli, G.~Cognola and D.~Klemm,
``Thermodynamics of Kerr-Newman-AdS black holes and conformal field theories,''
Class. Quant. Grav. \textbf{17}, 399-420 (2000)
doi:10.1088/0264-9381/17/2/310
[arXiv:hep-th/9908022 [hep-th]].
\bibitem{Kastor:2009wy}
D.~Kastor, S.~Ray and J.~Traschen,
``Enthalpy and the Mechanics of AdS Black Holes,''
Class. Quant. Grav. \textbf{26}, 195011 (2009)
doi:10.1088/0264-9381/26/19/195011
[arXiv:0904.2765 [hep-th]].




\bibitem{Dolan:2011xt}
B.~P.~Dolan,
``Pressure and volume in the first law of black hole thermodynamics,''
Class. Quant. Grav. \textbf{28}, 235017 (2011)
doi:10.1088/0264-9381/28/23/235017
[arXiv:1106.6260 [gr-qc]].

\bibitem{Dolan:2014jva}
B.~P.~Dolan,
``Black holes and Boyle's law \textemdash{} The thermodynamics of the cosmological constant,''
Mod. Phys. Lett. A \textbf{30}, no.03n04, 1540002 (2015)
doi:10.1142/S0217732315400027
[arXiv:1408.4023 [gr-qc]].





\bibitem{Huang:1995zb}
C.~G.~Huang and C.~B.~Liang,
``A Torus like black hole,''
Phys. Lett. A \textbf{201}, 27-32 (1995)
doi:10.1016/0375-9601(95)00229-V

\bibitem{Lemos:1994xp}
J.~P.~S.~Lemos,
``Cylindrical black hole in general relativity,''
Phys. Lett. B \textbf{353}, 46-51 (1995)
doi:10.1016/0370-2693(95)00533-Q
[arXiv:gr-qc/9404041 [gr-qc]].

\bibitem{Hong:2020zcf}
W.~Hong, B.~Mu and J.~Tao,
``Testing the weak cosmic censorship conjecture in torus-like black hole under charged scalar field,''
Int. J. Mod. Phys. D \textbf{29}, no.12, 2050078 (2020)
doi:10.1142/S0218271820500789
[arXiv:2001.09008 [physics.gen-ph]].

\bibitem{Lemos:1995cm}
J.~P.~S.~Lemos and V.~T.~Zanchin,
``Rotating charged black string and three-dimensional black holes,''
Phys. Rev. D \textbf{54}, 3840-3853 (1996)
doi:10.1103/PhysRevD.54.3840
[arXiv:hep-th/9511188 [hep-th]].

\bibitem{Han:2019kjr}
Y.~W.~Han, X.~X.~Zeng and Y.~Hong,
``Thermodynamics and weak cosmic censorship conjecture of the torus-like black hole,''
Eur. Phys. J. C \textbf{79}, no.3, 252 (2019)
doi:10.1140/epjc/s10052-019-6771-y
[arXiv:1901.10660 [hep-th]].

\bibitem{Feng:2021vey}
H.~Feng, Y.~Huang, W.~Hong and J.~Tao,
``Charged torus-like black holes as heat engines,''
Commun. Theor. Phys. \textbf{73}, no.4, 045403 (2021)
doi:10.1088/1572-9494/abe3ef
[arXiv:2102.04603 [gr-qc]].


		\bibitem{Banados:1992wn}
		M.~Banados, C.~Teitelboim and J.~Zanelli,
		``The Black hole in three-dimensional space-time,''
		Phys. Rev. Lett. \textbf{69}, 1849-1851 (1992)
		doi:10.1103/PhysRevLett.69.1849
		[arXiv:hep-th/9204099 [hep-th]].
		
		\bibitem{Banados:1992gq}
		M.~Banados, M.~Henneaux, C.~Teitelboim and J.~Zanelli,
		``Geometry of the (2+1) black hole,''
		Phys. Rev. D \textbf{48}, 1506-1525 (1993)
		[erratum: Phys. Rev. D \textbf{88}, 069902 (2013)]
		doi:10.1103/PhysRevD.48.1506
		[arXiv:gr-qc/9302012 [gr-qc]].
		
		\bibitem{Paul:2023qvh}
		P.~Paul and S.~I.~Kruglov,
		``Thermodynamics of BTZ Black Holes in Nonlinear Electrodynamics,''
		[arXiv:2302.05704 [gr-qc]].
		
		\bibitem{Mo:2018hav}
		J.~X.~Mo and S.~Q.~Lan,
		``Phase transition and heat engine efficiency of phantom AdS black holes,''
		Eur. Phys. J. C \textbf{78}, no.8, 666 (2018)
		doi:10.1140/epjc/s10052-018-6153-x
		[arXiv:1803.02491 [gr-qc]].
		
		\bibitem{Quevedo:2016cge}
		H.~Quevedo, M.~N.~Quevedo and A.~Sanchez,
		``Geometrothermodynamics of phantom AdS black holes,''
		Eur. Phys. J. C \textbf{76}, no.3, 110 (2016)
		doi:10.1140/epjc/s10052-016-3949-4
		[arXiv:1601.07120 [gr-qc]].
		
		\bibitem{Jardim:2012se}
		D.~F.~Jardim, M.~E.~Rodrigues and M.~J.~S.~Houndjo,
		``Thermodynamics of phantom Reissner-Nordstrom-AdS black hole,''
		Eur. Phys. J. Plus \textbf{127}, 123 (2012)
		doi:10.1140/epjp/i2012-12123-x
		[arXiv:1202.2830 [gr-qc]].
	\end{thebibliography}
\end{document}